\documentstyle[12pt,epsfig]{article}        
\newlength{\dinwidth}                       
\newlength{\dinmargin}                      
\def\lapprox{\lower .7ex\hbox{$\;\stackrel{\textstyle <}{\sim}\;$}}
\def\gapprox{\lower .7ex\hbox{$\;\stackrel{\textstyle >}{\sim}\;$}}

\def\d{{\rm d}}

\def\lsim{\mathrel{\rlap{\lower4pt\hbox{\hskip1pt$\sim$}}
    \raise1pt\hbox{$<$}}}                
\def\gsim{\mathrel{\rlap{\lower4pt\hbox{\hskip1pt$\sim$}}
    \raise1pt\hbox{$>$}}}                
\begin{document}
\begin{flushright}
DTP/96/84
\end{flushright}
\vspace*{-0.2cm}
\begin{center}  \begin{Large} \begin{bf}
Time-reversal-odd asymmetries at HERA\footnote{Contribution to the
proceedings of the ``Future Physics at HERA''-workshop,
DESY, Hamburg, 1995/96.}\\
  \end{bf}  \end{Large}
  \vspace*{5mm}
  \begin{large}
T.~Gehrmann$^a$\\ 
  \end{large}
\end{center}
$^a$ Department of Physics, 
University of Durham,
South Road, Durham, UK \\
\begin{quotation}
\noindent
{\bf Abstract:}
Longitudinal polarization of the HERA electron beam allows the study
of spatial asymmetries in the final state of single inclusive hadron
production in $\vec{e}p$-collisions. We estimate the size of a
particular production asymmetry which is related to the time reversal
properties of the underlying parton level subprocess.
\end{quotation}

Despite the experimental succes of QCD, some of its basic interaction 
properties are yet undetermined. In particular, only little
is known about the behaviour of QCD interactions under discrete
symmetry operations. 

It was shown in~\cite{ru} that time-reversal properties of a
particular interaction can be tested by measuring experimental
observables which are odd under the time-reversal operation. Assuming
invariance under $T$, these
$T$-odd observables vanish at tree level but receive contributions
from the absorptive part of higher order loop corrections, yielding in
general a non-zero effect. 

The first measurement of a $T$-odd QCD observable was
carried out in hadronic $Z^0$ decays
by the SLD collaboration~\cite{sld}. Due to the presence of $T$-odd
contributions from QCD and electroweak interactions~\cite{bb} at $\sqrt{s} =
M_Z$, this measurement fails to provide a clean probe of the
$T$-properties of QCD. 
A complimentary measurement of a $T$-odd observable in deep inelastic
scattering at moderate $Q^2$ would on the contrary be insensitive to
electroweak effects. 

Such an observable is suggested in~\cite{hag,mulders}: the azimuthal angle
between the momenta of the outgoing electron and an outgoing hadron in
the plane transverse to the current direction in the target rest
frame. This angle can be expressed as a triple product
\begin{displaymath}
\sin \phi = \frac{\vec{s}\times\vec{k}'\cdot \vec{P_T}'}{ 
|\vec{s}\times\vec{k}'|| \vec{P_T}'|}\; , 
\end{displaymath} 
where $\vec{s}$ denotes the incoming electron spin, $\vec{k}'$ 
 the momentum of the outgoing electron and  $\vec{P_T}'$ the
transverse momentum of the outgoing hadron relative to the current
direction.
\begin{figure}[t]
\begin{center}
~ \epsfig{file=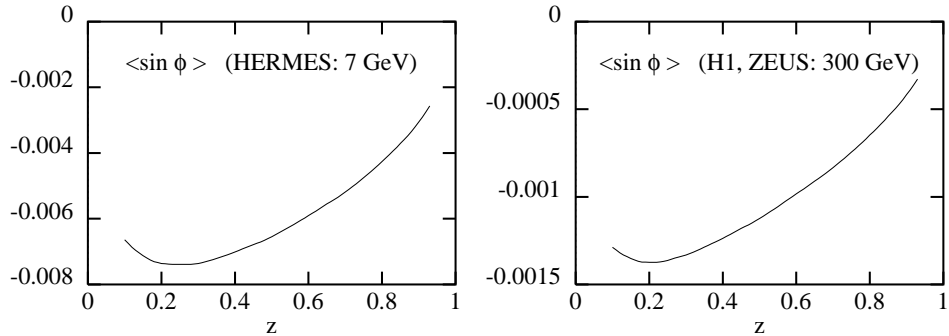,width=12.7cm}
\caption{{\it
Expected angular asymmetry in single inclusive hadron production
with a beam of left-handed electrons. 
  }}
\label{fig:todd}
\end{center}
\end{figure}

The leading non-zero contribution inducing a $\sin \phi$-dependence 
into the single hadron inclusive cross section 
arises from the absorptive part of the 
one-loop corrections to the
$\gamma^{\star} q \to q g$ and $\gamma^{\star}g \to q \bar q$
subprocesses and was calculated in~\cite{hag}. The average angle
$\langle \sin
\phi\rangle$ is
\begin{equation}
\langle\sin \phi\rangle (z) = \int \d x \d y \d \phi \; \sin
\phi\; \frac{\d \sigma}{\d x \d y \d \phi\d z}\; \Bigg/ \; \int \d x \d
y \d \phi\; \frac{\d \sigma}{\d x \d y \d \phi\d z} .
\label{eq:toddmaster}
\end{equation}
As the numerator is ${\cal O}(\alpha_s^2)$ while the denominator is
${\cal O}(1)$, one should expect $\langle \sin \phi\rangle$ to be rather
small. The non-vanishing of  $\langle \sin \phi\rangle$ is manifest as
asymmetry between the left and right event hemispheres in a frame
defined by the current direction and divided by the lepton scattering
plane. It has to be kept in mind that this asymmetry is subleading to
the up--down asymmetry in the same frame, which arises from a $\cos
\phi$ dependence of the cross section at ${\cal
O}(\alpha_s)$~\cite{updown}. 

In order to estimate the expected asymmetries at HERA and at HERMES, 
we
have evaluated the average angle $\langle \sin \phi\rangle$ 
in the single inclusive
production of charged pions and kaons, using the LO parton
distribution functions of~\cite{grvlo} and the LO hadron fragmentation
functions of~\cite{bkk}.  
 To suppress higher
twist contributions which can yield sizable $T$-odd
asymmetries~\cite{mulders}, we have restricted the final state to
$Q^2> 5\;\mbox{GeV}^2$ and  $W^2> 5\;\mbox{GeV}^2$. The resulting
average angle $\langle \sin \phi\rangle$ for left-handed electrons 
is shown in Fig.~\ref{fig:todd} as a
function of the longitudinal hadron momentum $z=(P\cdot P')/(P\cdot q)$.
The asymmetry for right-handed electrons would be equal in magnitude
but opposite in sign.

Although the asymmetry appears small at first sight, a measurement
might still be possible. The denominator of (\ref{eq:toddmaster}) can
be suppressed to ${\cal O}(\alpha_s)$ by including only hadrons above a
minimal transverse momentum in the measurement. A realistic estimate
with a transverse momentum cut has to take effects due to detector
resolution and geometry into account and is beyond the scope of the
present study.


\begin{thebibliography}{9}
\bibitem{ru}
A.~De R\'{u}jula, J.~M.~Kaplan and E.~de Rafael, Nucl.~Phys. {\bf B35}
(1971) 365.
\bibitem{sld}
SLD collaboration: K.~Abe et al., Phys.~Rev.~Lett. {\bf 75} (1995)
4173.
\bibitem{bb}
A.~Brandenburg, L.~Dixon and Y.~Shadmi, Phys.~Rev. {\bf D53} (1996) 1264.  
\bibitem{hag}
K.~Hagiwara, K.~Hikasa and N.~Kai, Phys.~Rev. {\bf D27} (1983) 84.
\bibitem{mulders}
J.~Levelt and P.~Mulders, Phys.~Lett. {\bf B338} (1994) 357.
\bibitem{updown}
J.~Chay, S.D.~Ellis and W.J.~Stirling, Phys.~Rev. {\bf D45} (1992) 46.
\bibitem{grvlo}
M.~Gl\"uck, E.~Reya and A.~Vogt, Z.~Phys. {\bf C67} (1995) 433.
\bibitem{bkk}
J.~Binnewies, B.A.~Kniehl and G.~Kramer, Phys.~Rev. {\bf D52} (1995)
4947.
\end{thebibliography}
\end{document}